\pdfoutput=1
\RequirePackage{ifpdf}
\ifpdf 
\documentclass[pdftex]{sigma}
\else
\documentclass{sigma}
\fi

\usepackage{bbm}

\newcommand{\norm}[1]{{\lVert #1 \rVert}}
\DeclareMathOperator{\Tr}{Tr}

\begin{document}

\allowdisplaybreaks

\renewcommand{\thefootnote}{$\star$}

\renewcommand{\PaperNumber}{067}

\FirstPageHeading

\ShortArticleName{An Index for Intersecting Branes in Matrix Models}

\ArticleName{An Index for Intersecting Branes in Matrix Models\footnote{This paper is a~contribution to the
Special Issue on Deformations of Space-Time and its Symmetries.
The full collection is available at \href{http://www.emis.de/journals/SIGMA/space-time.html}
{http://www.emis.de/journals/SIGMA/space-time.html}}}

\Author{Harold STEINACKER and Jochen ZAHN}

\AuthorNameForHeading{H.~Steinacker and J.~Zahn}

\Address{Fakult\"at f\"ur Physik, Universit\"at Wien, Boltzmanngasse 5, 1090 Wien, Austria}
\Email{\href{mailto:harold.steinacker@univie.ac.at}{harold.steinacker@univie.ac.at},
\href{mailto:jochen.zahn@univie.ac.at}{jochen.zahn@univie.ac.at}}

\ArticleDates{Received September 17, 2013; Published online November 08, 2013}

\vspace{-1mm}

\Abstract{We introduce an index indicating the occurrence of chiral fermions at the intersection of branes
in matrix models. This allows to discuss the stability of chiral fermions under perturbations of the branes.}


\Classification{81R60; 81T75; 81T30}

\renewcommand{\thefootnote}{\arabic{footnote}}
\setcounter{footnote}{0}

\vspace{-2mm}

\section{Introduction}

The IKKT or IIB model~\cite{IKKT} admits solutions which can be interpreted as branes embedded in a~f\/lat
target space, cf.~\cite{SteinackerEmergentGravity, SteinackerIntroduction} for recent reviews.
Of particular interest is the case of intersecting branes, as these can give rise to chiral fermions living
on the intersection~\cite{CSZ_IntersectingBranes}, thus having the potential of yielding a~physically
viable model.
The aim of this note is to show that the occurrence of chiral fermions can be rephrased as the
non-vanishing of a~certain index, which counts the number of zero modes of the Dirac operator, weighted
with their chirality.
Moreover, conditions ensuring the stability of the index under perturbations are given.
This is also demonstrated in a~concrete example.

Let us explain the relation of the index we propose with dif\/ferent indices discussed in the context of
emergent geometries in matrix models.
In~\cite{Berenstein}, two-dimensional compact branes embedded in ${\mathbb{R}}^3$ are studied.
In our language, the intersection of such branes with a~point $y \in {\mathbb{R}}^3$ is considered, and the
index counts the dif\/ference of the number of positive and negative modes of the corresponding Dirac
operator.
The set of $y$'s where the index changes is then interpreted as the locus of the brane.
Crucial dif\/ferences to our setting are the odd dimension of the target space (so that there is no
chirality operator and eigenvalues of the Dirac operator are not symmetric around $0$), and the restriction
to f\/inite matrices.

Another index for noncommutative branes was considered in~\cite{AokiChiralFermions}.
The dif\/ference to our def\/inition is the usage of another Dirac operator, the so-called Ginsparg--Wilson
Dirac operator, which does not coincide with the Dirac operator appearing in the IKKT action.

The article is structured as follows: In the next section, we recall aspects of the matrix model framework
and its ef\/fective (brane) geometry.
We introduce the notion of intersecting branes, and introduce the index indicating the occurrence of chiral
fermions.
In Section~\ref{sec:Deformation}, we present conditions guaranteeing the stability of the index
under deformations, and discuss a~concrete example.
We conclude with a~brief summary and an outlook.

\section{Matrix models, intersecting branes, and chiral fermions}

We brief\/ly collect the essential ingredients of the matrix model framework and its ef\/fective geometry,
referring to the recent review~\cite{SteinackerIntroduction} for more details.
The starting point is the maximally supersymmetric IKKT or IIB model~\cite{IKKT}, whose action is given by
\begin{gather*}
S=\Tr\left(\big[X^A,X^B\big]\big[X^C,X^D\big]\eta_{AC}\eta_{BD}+2\bar\Psi D\Psi\right).
\end{gather*}
Here the $X^A$ are Hermitian matrices, i.e., operators acting on a~separable Hilbert space $\mathcal{H}$.
The indices run from $0$ to $9$, and will be raised or lowered with the invariant tensor $\eta_{AB}$ of~${\rm SO}(9,1)$.
Furthermore, $\Psi$ is a~matrix-valued Majorana Weyl spinor of ${\rm SO}(9,1)$, and $D$ is the Dirac ope\-rator,
def\/ined by
\begin{gather*}
(D\Psi)^a={\Gamma_A}^a_b\big[X^A,\Psi^b\big],
\end{gather*}
where $\Gamma_A$ are the $10$-dimensional $\gamma$ matrices.

Even though this will not be used explicitly, the picture to have in mind is that the matrix
conf\/igurations $X^A$ describe embedded noncommutative branes.
By this one means that the~$X^A$ can be interpreted as quantized embedding
functions~\cite{SteinackerIntroduction} of a~$2n$-dimensional sub\-mani\-fold ${\mathcal{M}}^{2n}
\hookrightarrow {\mathbb{R}}^{10}$.
More precisely, there should be some quantization map ${\mathcal{Q}}: {\mathcal{C}}({\mathcal{M}}^{2n}) \to
L({\mathcal{H}})$ which maps classical functions on ${\mathcal{M}}^{2n}$ to a~noncommutative (matrix)
algebra, such that commutators can be interpreted as quantized Poisson brackets.
The $X^A$ are then the image of classical embedding functions $x^A$ under this map.
For more details, we refer to~\cite{SteinackerIntroduction}.

If the matrices $X^A$ are of block-diagonal form
\begin{gather*}
X^A=
\begin{pmatrix}
X^A_L&0
\\
0&X^A_R
\end{pmatrix}
,
\end{gather*}
we speak of two intersecting branes.
If we analogously split the fermions as
\begin{gather*}
\Psi=
\begin{pmatrix}
\Psi_{LL}&\Psi_{LR}
\\
\Psi_{RL}&\Psi_{RR}
\end{pmatrix}
,
\end{gather*}
then the Dirac operator acts on the of\/f-diagonal components as
\begin{gather*}
D_{LR}\Psi_{LR}=\Gamma_A\left(X^A_L\Psi_{LR}-\Psi_{LR}X^A_{R}\right).
\end{gather*}
We will consider the case when, in the semiclassical picture, the two branes ${\mathcal{M}}_{L/R}$ are of
the form
\begin{gather}
\label{eq:SplitM}
{\mathcal{M}}_{L/R}={\mathcal{M}}^d\times{\mathcal{M}}'_{L/R},
\end{gather}
where $d$ is even and ${\mathcal{M}}^d$ is embedded in the subspace of ${\mathbb{R}}^{10}$ generated by
$e^\mu$, $0 \leq \mu \leq d-1$ and ${\mathcal{M}}'_{L/R}$ are embedded in the directions spanned by $e^i$,
$d \leq i \leq 9$.
Furthermore, the symplectic form on ${\mathcal{M}}_{L/R}$ is required to respect the
split~\eqref{eq:SplitM}, i.e., it should vanish for one vector in $T {\mathcal{M}}^d$ and one in $T
{\mathcal{M}}'_{L/R}$.
In formal terms, this means that
\begin{gather}
\label{eq:SplitHilbert}
{\mathcal{H}}_{L/R}={\mathcal{H}}^{(d)}_{L/R}\otimes{\mathcal{H}}^{(10-d)}_{L/R},
\qquad
X^\mu_{L/R}=Y^\mu_{L/R}\otimes\mathbbm{1},
\qquad
X^i_{L/R}=\mathbbm{1}\otimes Y^i_{L/R}.
\end{gather}
Here $\mu$ labels the indices $0 \leq \mu \leq d-1$, whereas $i$ labels $d \leq i \leq 9$.
Furthermore,
\begin{gather}
\label{eq:CommonBrane}
{\mathcal{H}}^{(d)}_L\simeq{\mathcal{H}}^{(d)}_R\simeq{\mathcal{H}}^{(d)}
\end{gather}
and, under this isomorphism, $X^\mu_L = X^\mu_R$.
This encodes the requirement that the two branes share a~common $d$-dimensional brane.
Using the identif\/ication of ${\mathcal{H}}^{(d)}_L$ and ${\mathcal{H}}^{(d)}_R$, we may write the Dirac
operator as
\begin{gather*}
D_{LR}\Psi_{LR}=D^{(d)}_{LR}\Psi_{LR}+D^{(10-d)}_{LR}\Psi_{LR}
=\Gamma_\mu[X^\mu,\Psi_{LR}]+\Gamma_i\left(X^i_L\Psi_{LR}-\Psi_{LR}X^i_{R}\right).
\end{gather*}
We also split the chirality operator (note the dif\/ferent signs in $\chi^{(d)}$ and $\chi^{(10-d)}$
stemming from the signature $(-, +, \dots, +)$ of $\eta$),
\begin{gather*}
\chi=\chi^{(d)}\chi^{(10-d)},
\qquad
\chi^{(d)}=i^{-d/2+1}\Gamma_0\cdots\Gamma_{d-1},
\qquad
\chi^{(10-d)}=i^{-(10-d)/2}
\Gamma_d\cdots\Gamma_{9},
\end{gather*}
and remark that it fulf\/ills
\begin{gather*}
\big[\chi^{(d)},\chi^{(10-d)}\big]=0,
\qquad
\big(\chi^{(d)}\big)^2=1,
\qquad
\big(\chi^{(10-d)}\big)^2=1,
\end{gather*}
and
\begin{gather*}
\big\{\chi^{(d)},D^{(d)}_{LR}\big\}=0,
\qquad
\big\{\chi^{(10-d)},D^{(10-d)}_{LR}\big\}=0,
\\
\big[\chi^{(d)}, D^{(10-d)}_{LR}\big]=0,
\qquad
\big[\chi^{(10-d)},D^{(d)}_{LR}\big]=0.
\end{gather*}
We also note that the $\Gamma$ matrices may be represented as
\begin{gather}
\label{eq:SplitGamma}
\Gamma_\mu=\gamma_\mu\otimes\mathbbm{1}_{2^{5-d/2}},
\qquad
\Gamma_i=\gamma_{d+1}\otimes\delta_i,
\end{gather}
where the $\gamma_\mu$ form the $d$-dimensional Lorentzian Clif\/ford algebra, $\gamma_{d+1}$ is the
corresponding chirality operator, and the $\delta_i$ form the $(10-d)$-dimensional Euclidean Clif\/ford
algebra.

Given that the $X_{L/R}$ are represented on Hilbert spaces ${\mathcal{H}}_{L/R}$, the of\/f-diagonal
fermions are elements of ${\mathcal{H}}_{LR} = B({\mathcal{H}}_R, {\mathcal{H}}_L) \otimes
{\mathbb{C}}^{2^5}$.
Due to the split~\eqref{eq:SplitHilbert}, a~general ansatz for solutions of $D_{LR} \Psi_{LR} = 0$
is\footnote{Note that the condition~\eqref{eq:SplitHilbert} is crucial here.
In~\cite{NishimuraAsato}, the same ansatz for $\Psi_{LR}$ is used, but~\eqref{eq:SplitHilbert} is not
fulf\/illed.
Hence, in that work, the ansatz is not general enough to f\/ind all solutions of the Dirac equation.}
\begin{gather*}
\Psi_{LR}=\Psi_{LR}^{(d)}\otimes\Psi_{LR}^{(10-d)},
\qquad
\Psi_{LR}^{(d)}\in{\mathcal{H}}^{(d)}_{LR},
\qquad
\Psi_{LR}^{(10-d)}\in{\mathcal{H}}^{(10-d)}_{LR},
\end{gather*}
where we def\/ined
\begin{gather*}
{\mathcal{H}}^{(d)}_{LR}=B\big({\mathcal{H}}^{(d)}\big)\otimes{\mathbb{C}}^{2^{d/2}},
\qquad
{\mathcal{H}}^{(10-d)}_{LR}=B\big({\mathcal{H}}_R^{(10-d)},{\mathcal{H}}_L^{(10-d)}\big)\otimes{\mathbb{C}}
^{2^{5-d/2}}.
\end{gather*}
Here we used~\eqref{eq:CommonBrane} and the same factorization of the spinorial representation space as
in~\eqref{eq:SplitGamma}.
Using the operator norm, ${\mathcal{H}}_{LR}^{(10-d)}$ can be given the structure of a~Banach space.
Due to~\eqref{eq:SplitGamma}, have
\begin{gather*}
D^{(10-d)}_{LR}=\gamma_{d+1}\otimes\Delta^{(10-d)}_{LR},
\qquad
\chi^{(10-d)}=\mathbbm{1}_{(d/2)^2}
\otimes\theta^{(10-d)},
\end{gather*}
where $\Delta^{(10-d)}_{LR}$ and $\theta^{(10-d)}$ are anticommuting operators on
${\mathcal{H}}_{LR}^{(10-d)}$.
In particular, non-zero eigenvalues of $\Delta^{(10-d)}_{LR}$ come in pairs $\pm m$, which are interchanged
by $\theta^{(10-d)}$, and whose eigenvectors $v_{\pm m}$ may be combined to eigenvectors $v^\pm_m$ of
$\theta^{(10-d)}$ of opposite chirality.
It is then clear that given an eigenvector $\Psi_{LR}^{(10-d)}$ of $\Delta^{(10-d)}_{LR}$ with eigenvalue
$m$, the Dirac equation for $\Psi^{(d)}_{LR}$ becomes, cf.~\eqref{eq:SplitHilbert},
\begin{gather*}
\Gamma^\mu\big[Y_\mu,\Psi^{(d)}_{LR}\big]+m\gamma_{d+1}\Psi^{(d)}_{LR}=0,
\end{gather*}
which does not admit chiral solutions unless $m=0$.
Furthermore, given a~zero mode $\Psi_{LR}^{(10-d)}$, the chirality of $\Psi_{LR}^{(10-d)}$ w.r.t.\
$\theta^{(10-d)}$ determines $\chi^{(d)} \Psi_{LR}^{(d)}$, i.e., the $d$-dimensional chirality of~$\Psi_{LR}^{(d)}$, by the total chirality constraint $\chi \Psi_{LR} = \Psi_{LR}$.
Hence, a~$d$-dimensional chiral fermion requires a~zero eigenvector of~$\Delta^{(10-d)}_{LR}$ with no
corresponding eigenvector of opposite chirality\footnote{Otherwise, their combination will in general
acquire a~mass through quantum corrections.}. Note that if we have a~chiral fermion in the $LR$ sector, then
the Majorana condition ensures that the $RL$ sector contains the conjugate fermion with opposite chirality.

By our assumptions, $\Delta^{(10-d)}_{LR}$ is a~Dirac operator on the intersection of branes with
Riemannian signature, so we may expect it to have discrete spectrum (in Section~\ref{sec:Example},
this is shown to be the case in a~concrete example).
The above discussion then motivates the following def\/inition of an index for the Dirac operator
$\Delta_{LR}^{(10-d)}$:
\begin{gather*}
\Xi\big(\Delta_{LR}^{(10-d)}\big)
=\Tr_{{\mathcal{H}}_{LR}^{(10-d)}}\left(P_\Gamma\big(\big(\Delta_{LR}^{(10-d)}\big)^2\big)\theta^{(10-d)}\right).
\end{gather*}
Here $\Gamma$ is some closed curve that encircles the origin and does not intersect an eigenvalue
of $\big(\Delta_{LR}^{(10-d)}\big)^2$,
and $P_\Gamma\big(\big(\Delta_{LR}^{(10-d)}\big)^2\big)$ is the orthogonal projector on the
eigenspaces whose eigenvalues are encircled by $\Gamma$.
As discussed above, nonzero eigenvalues of $\big(\Delta_{LR}^{(10-d)}\big)^2$ occur in pairs of opposite
chirality, so the def\/inition is independent of the choice of $\Gamma$.
The index counts the number of 0 eigenmodes, weighted with their chirality.
This index can also be written in the form
\begin{gather*}
\Xi\big(\Delta_{LR}^{(10-d)}\big)
=\Tr_{{\mathcal{H}}_{LR}^{(10-d)}}\Big(e^{-t\big(\Delta_{LR}^{(10-d)}\big)^2}\theta^{(10-d)}\Big)
\end{gather*}
for generic $t > 0$, which is analogous to the usual def\/inition of the index on compact Riemannian
spaces, cf.~\cite[Theorem~3.50]{BerlineGetzlerVergne}.

The motivation for introducing an index to describe chirality is that it takes discrete values, so by
continuity, one would expect it to be constant under deformations of the branes.
In the next section, we will discuss criteria which indeed ensure this.

\section{Deformation stability of chiral modes}
\label{sec:Deformation}

Let us begin by recalling a~notion from perturbation theory.
Let $A$ be a~closed, in general unbounded operator on a~Banach space.
Then $B$ is $A$-bounded, if $D(A) \subset D(B)$, and there are positive constants $a$, $b$ such that
\begin{gather*}
\norm{B x}\leq a\norm{A x}+b\norm{x}
\end{gather*}
holds for all $x \in D(A)$.
A straightforward consequence of~\cite[Theorem~IV.3.18]{Kato} is now the following:
\begin{proposition}
Let $A$ have discrete spectrum.
Given a~closed curve $\Gamma$ in ${\mathbb{C}}$ that encircles a~finite part of the spectrum, we define
the projector $P_\Gamma(A)$ on the corresponding eigenspaces.
Given an $A$-bounded operator~$B$, the map $\lambda \mapsto P_\Gamma(A + \lambda B)$ is norm-continuous in
a~small enough neighborhood of~$0$.
\end{proposition}

Now f\/ix some $X^i_{L/R}$.
By the above proposition and the fact that $\Xi$ takes discrete values, one easily obtains precise
conditions that ensure the invariance of the index under perturbations of the $X^i_{L/R}$:
\begin{proposition}
Let $\tilde X^i_{L/R} \in L\big({\mathcal{H}}_{L/R}^{(10-d)}\big)$ be self-adjoint and $\tilde
\Delta^{(10-d)}_{LR}$ the corresponding Dirac operator.
Assume that $\big({\tilde \Delta}_{LR}^{(10-d)}\big)^2$
and $\big\{\tilde \Delta_{LR}^{(10-d)}, \Delta_{LR}^{(10-d)}\big\}$ are $\big(\Delta_{LR}^{(10-d)}\big)^2$ bounded.
Then there is a~neighborhood $U$ of $0$ such that
\begin{gather*}
\Xi\big(\Delta_{LR}^{(10-d)}+\lambda\tilde\Delta_{LR}^{(10-d)}\big)
=\Xi\big(\Delta_{LR}^{(10-d)}\big)
\end{gather*}
for all $\lambda \in U$.
\end{proposition}
\begin{remark}
\label{rem:FinDim}
If ${\mathcal{H}}_{L/R}^{(10-d)}$ are f\/inite-dimensional, then ${\mathcal{H}}_{LR}^{(10-d)}$ has f\/inite
even dimension.
It is then no longer necessary to restrict the trace to a~f\/inite number of eigenvalues, so one can
dispose of the projector in the def\/inition of $\Xi$.
It follows that for f\/inite-dimensional representation spaces (corresponding to compact branes), the
chirality index always vanishes.
\end{remark}

\subsection{An example}
\label{sec:Example}

Up to now, the discussion was generic, in particular independent of the commutation relations of the $X^i$.
Let us now consider the concrete example of intersecting Moyal planes (recall that a~Moyal plane is
def\/ined by canonical commutation relations $[X^i, X^j] = i \Theta^{ij}$, with $\Theta$ a~real
antisymmetric matrix).
Take $d=6$, and let $X^i_{L/R}$ span two 2-dimensional orthogonal Moyal planes, i.e.,
\begin{gather*}
X^6_L=x,
\qquad
X^7_L=p_x,
\qquad
X^8_L=0,
\qquad
X^9_L=0,
\\
X^6_R=0,
\qquad
X^7_R=0,
\qquad
X^8_R=y,
\qquad
X^9_R=p_y,
\end{gather*}
where $(x, p_x)$ and $(y, p_y)$ are the canonical position and momentum operators on
${\mathcal{H}}_{L/R}^{(10-d)} = L^2({\mathbb{R}})$.
As shown in~\cite{CSZ_IntersectingBranes} (and also below), the index of this conf\/iguration is $1$.
It is easy to see that
\begin{gather}
\label{eq:Delta2}
\big(\Delta_{LR}^{(10-d)}\big)^2=x^2+y^2+p_x^2+p_y^2+2\Sigma_{67}+2\Sigma_{89},
\end{gather}
where
\begin{gather*}
\Sigma_{ij}=\tfrac{i}{4}[\Gamma^i,\Gamma^j].
\end{gather*}
This operator acts on ${\mathcal{H}}^{(10-d)}_{LR} \simeq L^2({\mathbb{R}}^2) \otimes {\mathbb{C}}^4$,
where we use that
\begin{gather*}
B\big(L^2({\mathbb{R}}),L^2({\mathbb{R}})\big)\simeq L^2\big({\mathbb{R}}^2\big).
\end{gather*}

As the f\/irst four terms on the r.h.s.\  of~\eqref{eq:Delta2} form a~positive def\/inite quadratic form,
it follows from the above proposition that the index is invariant under perturbations $X^i_{L/R} \to
X^i_{L/R} + \lambda \tilde X^i_{L/R}$ for small enough $\lambda$, if the $\tilde X^i_{L/R}$ are bounded or
linear (corresponding to intersections at angles\footnote{For intersections at angles in the context of
string compactif\/ications, cf.~\cite{BerkoozDouglasLeigh, BlumenhagenCveticLangacker, Gauntlett97}.}) in
the $X^i_{L/R}$ (or a~sum of such contributions).
In order to see this explicitly, let us consider the case where the $\tilde X^i$ are linear in the $X^i$.
As an example, consider
\begin{gather*}
X^6=x+c y,
\qquad
X^7=p_x,
\qquad
X^8=y,
\qquad
X^9=p_y.
\end{gather*}
For the square of the Dirac operator, one obtains
\begin{gather*}
\big(\Delta_{LR}^{(10-d)}\big)^2=\underbrace{x^2+\big(1+c^2\big)y^2-2c xy+p_x^2+p_y^2}_{\Delta_1}
+\underbrace{2\Sigma_{67}-2\Sigma_{89}+2c\Sigma_{69}}_{\Delta_2}.
\end{gather*}
Here $\Delta_1$ acts on $L^2({\mathbb{R}}^2)$, while $\Delta_2$ acts on the spinorial representation space
${\mathbb{C}}^4$.
To have a~zero eigenvector of $\big(\Delta_{LR}^{(10-d)}\big)^2$ requires a~pair of eigenvectors of $\Delta_1$
and $\Delta_2$ which add up to zero.
Let us compute the lowest eigenvalue of $\Delta_1$.
We use the ansatz
\begin{gather*}
\Psi=e^{-\frac{1}{2}(A x^2+B y^2+2C xy)}.
\end{gather*}
The eigenvalue equation $\Delta_1 \Psi = \eta \Psi$ then leads to
\begin{gather*}
-A^2-C^2+1=0,
\\
-C^2-B^2+1+c^2=0,
\\
-AC-BC-c=0,
\end{gather*}
the eigenvalue being given by $\eta = A+B$.
It is straightforward to f\/ind the eigenvalue $\eta = \sqrt{4 + c^2}$.
For the eigenvalues of the spinorial part $\Delta_2$, one f\/inds
\begin{gather*}
\eta=\pm c,\qquad\eta=\pm\sqrt{4+c^2}.
\end{gather*}
Hence, there is exactly one way to cancel the eigenvector of $\Delta_1$, i.e., there is one eigenvector of
$\big(\Delta_{LR}^{(10-d)}\big)^2$ with eigenvalue $0$ (the higher eigenvalues of $\Delta_1$ can obviously not
lead to further zero eigenvalues).
One can also explicitly check that it has positive chirality.
Analogously, one can treat the $d=4$ dimensional intersection of a~$6$- and an $8$-dimensional brane, and
similar conf\/igurations~\cite{CSZ_IntersectingBranes}.

An example of intersecting branes with a~vanishing index is provided by a~degenerate intersection of two
quantum planes, such as
\begin{gather*}
X^6=x+y,
\qquad
X^7=p_x,
\qquad
X^8=0,
\qquad
X^9=p_y.
\end{gather*}
In this case the part of $\big(\Delta_{LR}^{(10-d)}\big)^2$ that is quadratic in the coordinates of the quantum
plane is given by
\begin{gather*}
(x-y)^2+p_x^2+p_y^2,
\end{gather*}
which is not a~positive def\/inite quadratic form.
In particular, the condition of being $\big(\Delta_{LR}^{(10-d)}\big)^2$ bounded is not fulf\/illed for rotations
of this plane.
One easily checks that the index for this conf\/iguration vanishes.
This underlines the necessity of spanning the full ${\mathbb{R}}^{(10-d)}$ in order to get chiral fermions,
as already pointed out in~\cite{CSZ_IntersectingBranes}.

\section{Summary and outlook}

We presented a~def\/inition of an index describing the occurrence of chiral fermions on intersecting branes
in matrix models and discussed the stability of this index under perturbations.
In particu\-lar, this implies the existence of chiral fermions for branes intersecting at angles.
The drawback of our approach is that it requires strong restrictions on the embedding, in
particular~\eqref{eq:SplitHilbert}.
It is for example not applicable for situations in which (in the semiclassical picture) the brane
${\mathcal{M}}^d$ is not f\/lat.
One possibility to treat this case could be to work in the semiclassical limit, or to use a~modif\/ied
chirality operator, like\footnote{This particular operator has the disadvantage that it does in general not
anticommute with the Dirac operator, but it may be useful nevertheless.}
\begin{gather*}
\chi={\varepsilon}^{A_1\dots A_{2n}C_1\dots C_{10-2n}}{\varepsilon}_{B_1\dots B_{2n}C_1\dots C_{10-2n}}
X^{B_1}\cdots X^{B_{2n}}\Gamma_{A_1}\cdots\Gamma_{A_{2n}}
\end{gather*}
for a~$2n$-dimensional brane.
We plan to come back to this issue in future work.

As noted in Remark~\ref{rem:FinDim}, the index always vanishes for intersections of compact fuzzy spaces
${\cal K}_i \subset {\mathbb{R}}^{(10-d)}$.
This raises an apparent paradox, since the results on chiral fermions on intersections should apply at
least approximately for each intersection.
What happens is that pairs of ``almost-localized'' fermionic near-zero modes arise on the intersections
${\mathcal{K}}_i \cap {\mathcal{K}}_j$, such that for each ``ef\/fectively'' chiral fermion localized on
some intersection, there is another fermion with opposite chirality at some other
intersection\footnote{This is verif\/ied in numerical simulations.}. This means that if, e.g., the chiral
fermions of the standard model arise from some intersections such as in~\cite{CSZ_IntersectingBranes},
there are additional sectors with fermions of opposite chirality localized at dif\/ferent intersections.
The approximate localization on dif\/ferent intersections suggests that these unwanted sectors could be
ef\/fectively hidden or removed in some way.
A natural strategy to achieve this is to give up the product ansatz~\eqref{eq:SplitHilbert}, as proposed
in~\cite{NishimuraAsato}, and as realized, e.g., by solutions with split
noncommutativity~\cite{SteinackerSplit}.
These are interesting directions for further research.

\subsection*{Acknowledgments}
This work was supported by the Austrian Science Fund (FWF) under the contract
P24713.

\pdfbookmark[1]{References}{ref}
\LastPageEnding


\begin{thebibliography}{99}
\footnotesize\itemsep=0pt

\bibitem{AokiChiralFermions}
Aoki H., Chiral fermions and the standard model from the matrix model
  compactif\/ied on a torus, \href{http://dx.doi.org/10.1143/PTP.125.521}{\textit{Progr. Theoret. Phys.}} \textbf{125} (2011),
  521--536, \href{http://arxiv.org/abs/1011.1015}{arXiv:1011.1015}.

\bibitem{Berenstein}
Berenstein D., Dzienkowski E., Matrix embeddings on f\/lat $R^3$ and the geometry
  of membranes, \href{http://dx.doi.org/10.1103/PhysRevD.86.086001}{\textit{Phys. Rev.~D}} \textbf{86} (2012), 086001, 19~pages,
  \href{http://arxiv.org/abs/1204.2788}{arXiv:1204.2788}.

\bibitem{BerkoozDouglasLeigh}
Berkooz M., Douglas M.R., Leigh R.G., Branes intersecting at angles,
  \href{http://dx.doi.org/10.1016/S0550-3213(96)00452-X}{\textit{Nuclear Phys.~B}} \textbf{480} (1996), 265--278,
  \href{http://arxiv.org/abs/hep-th/9606139}{hep-th/9606139}.

\bibitem{BerlineGetzlerVergne}
Berline N., Getzler E., Vergne M., Heat kernels and {D}irac operators,
  \textit{Grundlehren der Mathematischen Wissenschaften}, Vol.~298,
  Springer-Verlag, Berlin, 1992.

\bibitem{BlumenhagenCveticLangacker}
Blumenhagen R., Cvetic M., Langacker P., Shiu G., Toward realistic intersecting
  D-brane models, \href{http://dx.doi.org/10.1146/annurev.nucl.55.090704.151541}{\textit{Ann. Rev. Nucl. Part. Sci.}} \textbf{55} (2005),
  71--139, \href{http://arxiv.org/abs/hep-th/0502005}{hep-th/0502005}.

\bibitem{CSZ_IntersectingBranes}
Chatzistavrakidis A., Steinacker H., Zoupanos G., Intersecting branes and a
  standard model realization in matrix models, \href{http://dx.doi.org/10.1007/JHEP09(2011)115}{\textit{J.~High Energy Phys.}}
  \textbf{2011} (2011), no.~9, 115, 36~pages, \href{http://arxiv.org/abs/1107.0265}{arXiv:1107.0265}.

\bibitem{Gauntlett97}
Gauntlett J.P., Intersecting branes, \href{http://arxiv.org/abs/hep-th/9705011}{hep-th/9705011}.

\bibitem{IKKT}
Ishibashi N., Kawai H., Kitazawa Y., Tsuchiya A., A large-{$N$} reduced model
  as superstring, \href{http://dx.doi.org/10.1016/S0550-3213(97)00290-3}{\textit{Nuclear Phys.~B}} \textbf{498} (1997), 467--491,
  \href{http://arxiv.org/abs/hep-th/9612115}{hep-th/9612115}.

\bibitem{Kato}
Kato T., Perturbation theory for linear operators, \textit{Die Grundlehren der
  mathematischen Wissenschaften}, Vol.~132, Springer-Verlag, New York, 1966.

\bibitem{NishimuraAsato}
Nishimura J., Tsuchiya A., Realizing chiral fermions in the type IIB matrix
  model at f\/inite~$N$, \href{http://arxiv.org/abs/1305.5547}{arXiv:1305.5547}.

\bibitem{SteinackerEmergentGravity}
Steinacker H., Emergent gravity from noncommutative gauge theory,
  \href{http://dx.doi.org/10.1088/1126-6708/2007/12/049}{\textit{J.~High Energy Phys.}} \textbf{2007} (2007), no.~12, 049, 36~pages,
  \href{http://arxiv.org/abs/0708.2426}{arXiv:0708.2426}.

\bibitem{SteinackerIntroduction}
Steinacker H., Emergent geometry and gravity from matrix models: an
  introduction, \href{http://dx.doi.org/10.1088/0264-9381/27/13/133001}{\textit{Classical Quantum Gravity}} \textbf{27} (2010), 133001,
  46~pages, \href{http://arxiv.org/abs/1003.4134}{arXiv:1003.4134}.

\bibitem{SteinackerSplit}
Steinacker H., Split noncommutativity and compactif\/ied brane solutions in
  matrix models, \href{http://dx.doi.org/10.1143/PTP.126.613}{\textit{Progr. Theoret. Phys.}} \textbf{126} (2011), 613--636,
  \href{http://arxiv.org/abs/1106.6153}{arXiv:1106.6153}.

\end{thebibliography}
\end{document}